\documentclass[aps,superscriptaddress,prb,preprint,double-spaced]{revtex4-2}

\usepackage{graphicx}
\usepackage{dcolumn}
\usepackage{bm}
\usepackage{amssymb,amsthm}
\usepackage{lineno}
\usepackage{subfigure}
\usepackage{textcomp}
\usepackage{booktabs}
\usepackage{xcolor}

\begin{document}

\title{High-temperature multigap superconductivity in two-dimensional metal-borides}

\author{Cem Sevik}
\thanks{These two authors contributed equally}
\affiliation{Department of Physics \& NANOlab Center of Excellence, University of Antwerp, Groenenborgerlaan 171, B-2020 Antwerp, Belgium}
\affiliation{Department of Mechanical Engineering, Faculty of Engineering, Eskisehir Technical University, 26555 Eskisehir, Turkey}
\author{Mikhail Petrov}
\thanks{These two authors contributed equally}
\affiliation{Department of Physics \& NANOlab Center of Excellence, University of Antwerp, Groenenborgerlaan 171, B-2020 Antwerp, Belgium}
\author{Jonas Bekaert}
\email{jonas.bekaert@uantwerpen.be}
\affiliation{Department of Physics \& NANOlab Center of Excellence, University of Antwerp, Groenenborgerlaan 171, B-2020 Antwerp, Belgium}
\author{Milorad V. Milo\v{s}evi\'{c}}
\email{milorad.milosevic@uantwerpen.be}
\affiliation{Department of Physics \& NANOlab Center of Excellence, University of Antwerp, Groenenborgerlaan 171, B-2020 Antwerp, Belgium}
\date{\today}

\begin{abstract}
Using first-principles calculations in combination with the Eliashberg formalism, we systematically investigated phonon-mediated superconductivity in two-dimensional (2D) metal-boride crystals, consisting of a boron honeycomb network doped by diverse metal elements. Such 2D metal-boride compounds, named MBenes, are chemically exfoliable from single-crystalline layered ternary borides (MAB phases). First we identified the MBene layers with potential for superconductivity via isotropic Eliashberg calculations, considering a wide range of metal elements, with focus on alkaline earth and transition metals. Subsequently, we performed a detailed analysis of the prominent superconducting MBenes by solving the anisotropic Eliashberg equations. The obtained high critical temperatures (up to 72 K), as well as the rich multigap superconducting behavior, recommend these crystals for further use in multifunctional 2D heterostructures and superconducting device applications.
\end{abstract}

\maketitle

\section{\label{sec:level1}Introduction}
The potential of two-dimensional (2D) materials in terms of future technological applications is well established~\cite{2d-one,2d-two, 2d-three}, as demonstrated for example through both theoretical and experimental studies with regard to transistors~\cite{transis1,transis2}, energy harvesters~\cite{2d-energy1, 2d-energy2, 2d-energy3, 2d-energy4}, and sensors~\cite{sensor1, sensor2, sensor3}. Recently, with the successful fabrication of 2D metallic materials such as In~\cite{In-SC}, Pb~\cite{Pb-SC}, NbSe$_2$~\cite{Nb-SC1, Nb-SC2}, FeSe~\cite{FeSe}, MgB$_2$~\cite{MgB-FewLayer}, Mo$_2$C~\cite{Mxenes2}, Nb$_2$C~\cite{Mxenes1}, doped graphene~\cite{graphene1, graphene2, graphene3}, and twisted bilayer graphene~\cite{graphene-twist}, superconductivity has been added to the range of properties yielding possible applications. However, taking advantage of that potential requires a detailed investigation and understanding of the origins and behavior of the superconductivity in these two-dimensional systems, to ensure reproducibly robust superconducting behavior up to sufficiently high temperatures. 

One of the well-known material families to examine the behavior of superconductivity at the 3D-to-2D transition are the layered hexagonal metal borides (MB$_2$) with \textit{P6/mmm} space group symmetry. Among them, bulk MgB$_2$ has been extensively studied over the past two decades, and is still the electron-phonon (\textit{e-ph}) mediated superconductor with highest experimentally realized superconducting transition temperature at ambient pressure to date ($T_c$ = 39~K)~\cite{mgb2-exp}. The experimental demonstration of two-gap superconductivity in this material has led to an immensely intensified interest in this area and a paradigm shift in superconductivity research. On that wave, superconductivity has also been experimentally validated in identical bulk crystals of ZrB$_2$~\cite{ZrB2-Exp}, NbB$_2$~\cite{NbB2-Exp}, and TaB$_2$~\cite{TaB2-Exp}. When thinned down to monolayer thicknesses, MgB$_2$ was demonstrated to harbor a very strong influence of the surface states \cite{jonas2}, leading to three-gap superconductivity with $T_c$ = 20~K in the monolayer limit ~\cite{jonas1}, strongly changing with every added monolayer, up to several nanometers thickness. Moreover, a strong increase in the $T_c$ of 2D MgB$_2$ (to over 100~K) with hydrogenation and tensile strain have been revealed~\cite{PhysRevLett.123.077001}. With the same crystal structure, single-layer AlB$_2$ has been predicted as a two-gap superconductor with $T_c$ = 26.5~K~\cite{AlB2-Calc, PhysRevB.101.104507}, surpassing single-layer MgB$_2$, although its bulk form is not superconducting at all. Last but not least, one is also interested in possible growth of MB$_{4}$ two-dimensional crystals, with the metal layer sandwiched between two boron honeycomb networks. In that respect, Li, Be, Mg, Al, and Ga-based MB$_{4}$ structures have been recently predicted to possess a multigap superconducting nature, all with $T_c$ above 30~K~\cite{PhysRevB.101.104507}. All together, the research to date clearly indicates the potential of layered metal boride systems to realize two-dimensional superconductivity. 

It is therefore instructive to have a comparative analysis of different possible 2D metal-borides. When considering the experimentally grown bulk metal-borides with the same crystal symmetry~\cite{https://doi.org/10.1002/adma.201604506}, one notices that \textit{e-ph} mediated superconducting properties of Ca, Sc, Ti, Zr, Hf, V, Nb, Ta, Cr, Mo, W and Re based 2D MBenes have not been explored. As a matter of fact, a recent first-principles study indicated the feasibility of synthesizing the layered MAB bulk crystals of hexagonal metal borides of Sc, Ti, Zr, Hf, V, Nb, Ta, Mo, and W, starting from the MAB bulk crystals \cite{C9NR01267B}, from which 2D MBenes are exfoliable (with name derived analogously to MXenes obtained from MAX bulk phases \cite{acsnano9b06394}). Note however that layer-by-layer exfoliation of these materials is not as straightforward as for graphene, due to the strong interaction between the metal and the boron layers. Therefore, one must consider several possible 2D formations, including MgB$_4$ (a metal layer sandwiched between two honeycomb B lattices, where the metal atoms are placed in the centers of honeycombs), MB$_2$ (a B layer in a honeycomb lattice and a layer of metal atoms sitting above the centers of honeycombs), and  M$_{2}$B$_2$ (a B layer in a honeycomb lattice sandwiched between two layers of metal atoms sitting above/below the centers of honeycombs). Therefore, in this paper we comprehensively investigated superconductivity in different exfoliable MBene structures, with particular attention to Ca, Sc, Ti, Zr, Hf, V, Nb, Ta, Cr, Mo, W, and Re based ones, in comparison to the previously studied Be, Mg, and Al based di- and tetra-borides. Following the selection of the most prominent superconducting MBenes, we detail their anisotropic superconducting properties, origin of their multiple superconducting gaps and their high critical temperature.

The organization of the paper is as follows. First, we describe the used computational approaches and adopted parameters for the calculations. Then, we present detailed isotropic Eliashberg theory results for all the considered materials. These results are then used to select the materials that potentially possess good superconducting properties. Finally, we summarize the results of detailed anisotropic Eliashberg calculations for the selected materials. Extensive additional results for electronic, vibrational, and superconducting properties are made available in the Supplementary Material.

\section{\label{sec:level2} Computational methods and details}
The considered metal boride layers, 45 in total, were investigated through density functional theory (DFT), as implemented within the ABINIT~\cite{Gonze2020, Gonze2016} code, using the Perdew-Burke-Ernzerhof (PBE) functional. Relativistic pseudopotentials from the PseudoDojo project~\cite{VANSETTEN201839} were used where we could include spin-orbit coupling (SOC) if necessary. Namely, we calculated electronic structures both without and with SOC for all compounds, and if significant differences were found in the states responsible for collective electron behavior, i.e. those near the Fermi level ($E_F$), we proceeded with the inclusion of SOC in further calculations. Within the used pseudopotentials $10+N$ valence electrons were used for metal elements of group $N$, except for Be (4 valence electrons), and 3 valence electrons were taken into account for B. An energy cutoff value of 60~Ha for the plane-wave basis and a dense 36$\times$36$\times$1 $k$-point grid were used to achieve high accuracy. The structures, with at least 16~{\AA} of vacuum, were relaxed so forces on each atom were below 1~meV/{\AA}. Subsequently, to calculate phonons and the electron–phonon coupling we employed density functional perturbation theory (DFPT), also within ABINIT~\cite{PhysRevB.54.16487}, using the same electronic $k$-point grid and a 12$\times$12$\times$1 phononic $q$-point grid. For the calculations of the superconducting state we used isotropic Eliashberg theory, a quantitatively accurate extension to the Bardeen–Cooper–Schrieffer (BCS) theory for phonon-mediated superconductivity~\cite{Eliashberg1, Eliashberg2, Eliashberg3}. We evaluated the superconducting $T_c$ using the Allen–Dynes formula \cite{PhysRev.167.331, PhysRevB.12.905, ALLEN19831}, and a screened Coulomb repulsion of $\mu^*$ = 0.13 -- a standard value for transition metal-based compounds~\cite{Grimvall}.
In addition, the cohesive energy values of the structures (per atom) were calculated using the following formula:
\begin{equation}
     E_{\mathrm{coh}} = \frac{{nE_M}+{lE_B}-E_{M_{n}B_{l}}}{n+l},
     \label{eq1}
 \end{equation}
where $E_M$ and $E_B$ are the total energies of the isolated metal and B atoms, respectively, and $E_{M_{n}B_{l}}$ is the total energy of the corresponding 2D MBene. In view of this expression, materials with more positive $E_{\mathrm{coh}}$ indicate a higher chemical stability.

Subsequently, for the materials predicted to possess high $T_c$ values, self-consistent solution of the fully anisotropic Migdal-Eliashberg equation was obtained through the Electron-Phonon Wannier (EPW) code~\cite{PhysRevB.76.165108,RevModPhys.89.015003,NOFFSINGER20102140}. For that part, first-principles calculations were performed with the Quantum ESPRESSO package~\cite{Giannozzi2009}, using the same PBE pseudopotentials. The cutoffs for the plane-wave basis and charge density were set to 100~Ry and 400~Ry, respectively. The $k$-point mesh for the integration in wave-vector space was set to 24$\times$24$\times$1, and electronic smearing was performed within the Methfessel-Paxton scheme~\cite{PhysRevB.40.3616}. With these parameters, the structures were fully optimized until the Hellman-Feynman forces on each atom were below $10^{-4}$~Ry/bohr. The subsequent dynamical properties were calculated based on the density-functional perturbation theory as implemented in  Quantum ESPRESSO package~\cite{RevModPhys.73.515}, using a $q$-mesh of 12$\times$12$\times$1. The maximally localized Wannier functions~\cite{PhysRevB.87.024505,MOSTOFI2008685} required in the EPW calculations were interpolated by using an unshifted Brillouin-zone $k$-mesh of 12$\times$12$\times$1. The solution of the anisotropic Migdal-Eliashberg equations was computed with interpolated $k$- and $q$-point grids of 240$\times$240$\times$1 and 120$\times$120$\times$1, which are sufficiently dense to guarantee convergence of the superconducting gaps. The cutoff for the fermion Matsubara frequencies $\omega_j = (2 j +1)\pi T$ was set to 0.5~eV and $\mu^*$ was the same as in the isotropic calculations (0.13).    

\begin {table}[h!]
\caption{Calculated lattice constant ($a$), electronic DOS at the Fermi level ($N_{F}$), average Fermi velocity ($v_F$), $e$-$ph$ coupling ($\lambda$), logarithmic average of phonon frequencies ($\omega_{ln}$), and Allen-Dynes superconducting transition temperature ($T_c$) for all the considered MBenes.}\label{table_l1} 
\begin{center}
\begin{tabular}{lcccccrr}
 Metal& $a$ & $N_F$ & $v_F$ & $\lambda$ & $\omega_{ln}$ & $T_c$\\
 &(\AA) & (eV$^{-1}$uc$^{-1}$)& (10$^{6}$ms$^{-1}$) & &  (K) & (K) \\\hline
 \multicolumn{7}{ c }{MB$_4$ monolayers}\\\hline
Be & 2.968 & 0.555 &  0.466 &  1.212 &  371 & 29.9\\
Mg & 3.007 & 0.716 &  0.678 &  0.808 &  573 & 22.2\\
Ca & 3.075 & 1.336 &  0.484 &  1.196 &  457 & 36.1\\
Sc & 3.080 & 1.100 &  0.324 &  0.714 &  371 & 10.4\\
Al & 2.991 & 0.996 &  0.527 &  0.911 &  615 & 30.9\\\hline
 \multicolumn{7}{ c }{MB$_2$ monolayers}\\\hline
Mg & 3.044 & 0.973 & 0.606 & 0.65 & 555 & 11.6\\
Ca & 3.220 & 2.523 & 0.277 & 1.67 & 360 & 41.6\\
Sc & 3.176 & 3.332 & 0.201 & 1.06 & 308 & 20.4\\
Zr & 3.152 & 1.455 & 0.211 & 0.67 & 125 & 2.9\\
V & 3.064 & 2.661 & 0.219 & 2.70 & 50 & 8.3\\
Nb & 2.997 & 3.436 & 0.191 & 2.23 & 245 & 35.5\\
Ta & 3.002  & 2.153 & 0.318 & 0.75 & 220 & 7.1\\
Cr & 3.128 & 3.302 & 0.311 & 1.68 & 36 & 4.5\\	
Re & 2.882 & 1.252 & 0.408 & 0.55 & 215 & 2.4\\
Al & 2.981 & 1.186 & 0.680 & 1.40 & 98 & 9.8\\\hline
 \multicolumn{7}{ c }{M$_2$B$_2$ monolayers}\\\hline
Mg & 3.104 & 0.749 &  0.665 &  0.486 &  498 & 3.2\\
Re & 2.916 & 1.285 &  0.317 &  1.088 &  81 & 5.5\\\hline
\end{tabular}
\end{center}
\end {table}

\section{\label{sec:level3}Results}
As the first step, the equilibrium lattice structures of the considered three different metal-boride structures were predicted and compared with the available data. The calculated in-plane lattice constants listed in the Supplementary Material are in good agreement both with available experimental values for bulk structures~\cite{ZrB2-Exp,MgB-FewLayer,PhysRevB.65.024303,https://doi.org/10.1002/adma.201604506} and with calculated values for the same single-layer crystals in the literature~\cite{AlB2-Calc,jonas2,mgb2-stabil}. The phonon dispersions presented in the Supplementary Material clearly point out the dynamical stability of all the considered crystals except BeB$_2$, TiB$_4$, ZrB$_4$, HfB$_4$, and ReB$_4$. Furthermore, the cohesive energy value, a widely accepted parameter used to evaluate the chemical stability of crystals (see Eq.~\ref{eq1}), was evaluated for each monolayer. Promisingly, the corresponding values presented in Table~\ref{table_CE} are comparable with the ones obtained for several experimentally realized monolayers, e.g., graphene (7.46 eV per atom), silicene (3.71 eV per atom), and black phosphorene (2.57 eV per atom). Based on these results, MB$_{4}$ is identified as the ground-state structure for Be, Mg, Ca, Sc, V, Cr, Mo, and Al based monolayers, while M$_2$B$_2$ was the favorable structure for the Ti, Zr, Hf, Nb, Ta, W, and Re based ones. 

\begin {table}[!ht]
\caption{Calculated cohesive energy, $E_{\mathrm{coh}}$, of the considered monolayer metal-boride structures.}\label{table_CE} 
\begin{center}
\begin{tabular}{lcccclccc}
& \multicolumn{3}{ c }{$E_{\mathrm{coh}}$ (eV/atom)} & & & \multicolumn{3}{ c }{$E_{\mathrm{coh}}$ (eV/atom)} \\\cline{2-4}\cline{7-9}
Metal & MB$_4$ & MB$_2$ & M$_2$B$_2$ & & Metal & MB$_4$ & MB$_2$ & M$_2$B$_2$ 
\\\hline
Be & 5.32 & 5.15 & 4.51 & & V & 6.04 & 5.58 & 5.76\\
Mg & 4.86 & 4.20 & 3.75 & & Nb & 6.40 & 6.14 & 6.68\\
Ca & 4.83 & 4.17 & 4.00 & & Ta & 6.64 & 6.48 & 7.16\\
Sc & 5.70 & 5.33 & 5.56 & & Cr & 5.60 & 4.94 & 4.75\\
Ti &  6.38 & 6.22 & 6.77 & & Mo & 6.15 & 5.73 & 5.95\\
Zr &  6.38 & 6.29 & 7.04 & & W & 6.55 & 6.35 & 6.75\\
Hf &  6.41 & 6.28 & 6.99 & & Re & 6.27 & 6.10 & 6.30\\
Al & 5.27 & 4.73 & 4.42 & & & & & \\\hline
\end{tabular}
\end{center}
\end {table}

Subsequently, we investigated the electronic properties of each material and we found nearly all the structures to be metallic. However, TiB$_2$ has a graphene-like band structure possessing a Dirac cone at the Fermi level as previously reported~\cite{mgb2-stabil}. Also, BeB$_2$, ZrB$_2$, HfB$_2$, CrB$_4$, MoB$_4$, and WB$_4$ were predicted to possess a very low electronic density of states at Fermi level due to single or multiple Dirac cones close to Fermi level, which is a clear obstacle for the formation of the superconducting state. As a result of this preliminary analysis, we eliminated these materials and computed the superconducting properties for the other candidates. Our isotropic Eliashberg theory calculations show that five MB$_4$, ten MB$_2$, and two M$_2$B$_2$ crystals from Table~\ref{table_l1} possess a sizable superconducting $T_c$. Among these materials, BeB$_4$, MgB$_4$, AlB$_4$, AlB$_2$, and MgB$_2$ have been previously investigated \cite{PhysRevB.101.104507,AlB2-Calc}. Indeed, our results for the superconducting transition temperature $T_c$ are comparable with the ones obtained in these references (where anisotropic Eliashberg theory was used). Also, the $T_c$ obtained for MgB$_4$ in this study (22.2~K), is in very good agreement with the one calculated by Liao \textit{et al.} in Ref.~\cite{C7CP06180C} (23.2~K).   

\begin{figure*}[!ht]
\includegraphics[width=\linewidth]{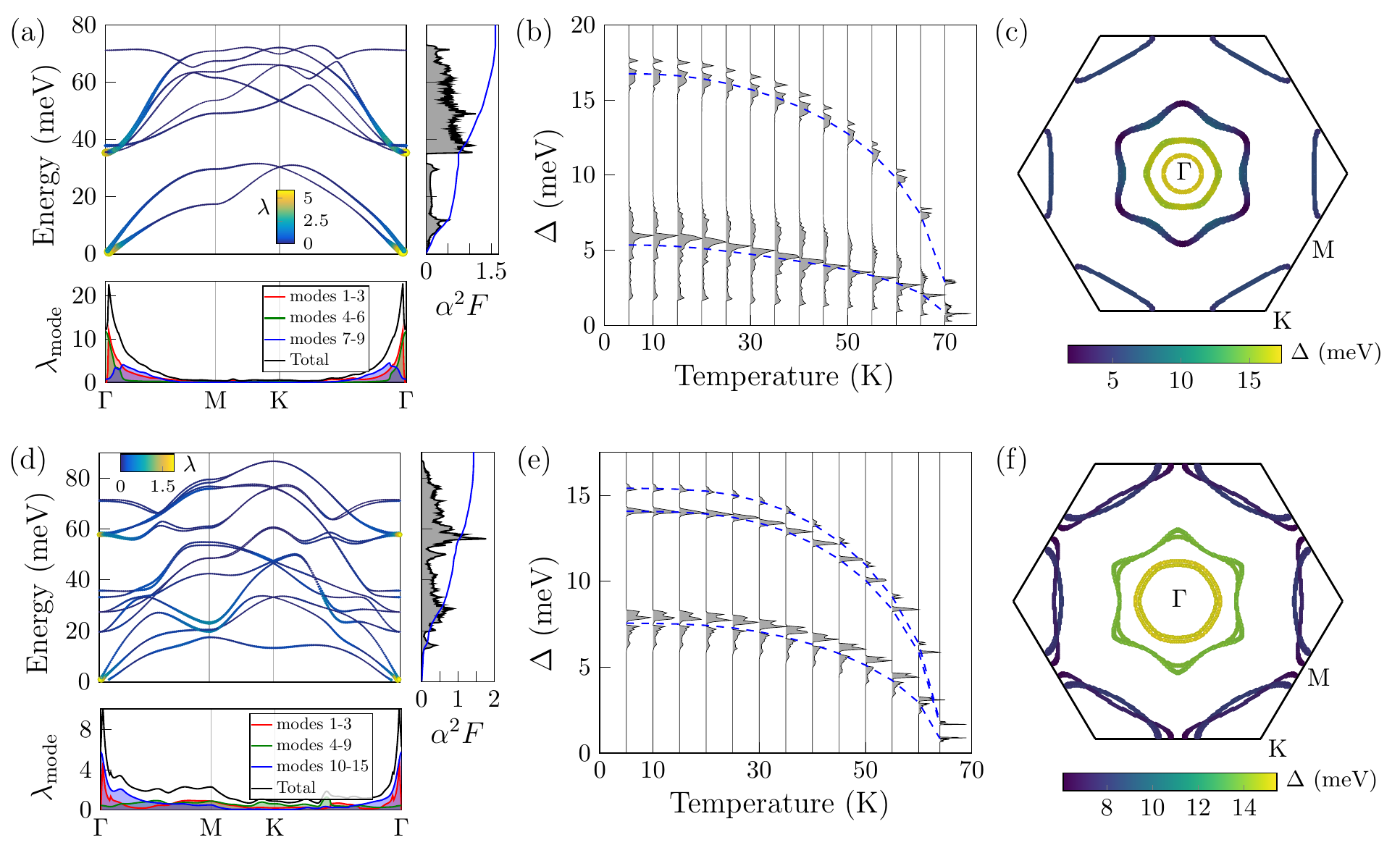}
\caption{ Phonon dispersion with the mode-resolved electron-phonon coupling (EPC) indicated by colors and thickness, EPC $\lambda_{\mathrm{mode}}$ summed for three different groups of phonon modes, isotropic Eliashberg function $\alpha^2F$, and EPC function $\lambda$ for (a) CaB$_2$ and (d) CaB$_4$. Evolution of the superconducting gap distribution for (b) CaB$_2$ and (e) CaB$_4$. Momentum-dependent superconducting gap on the Fermi surface at 5 K for (c) CaB$_2$ and (f) CaB$_4$.}
\label{fig1}
\end{figure*}

\begin{figure*}[!ht]
\includegraphics[width=\linewidth]{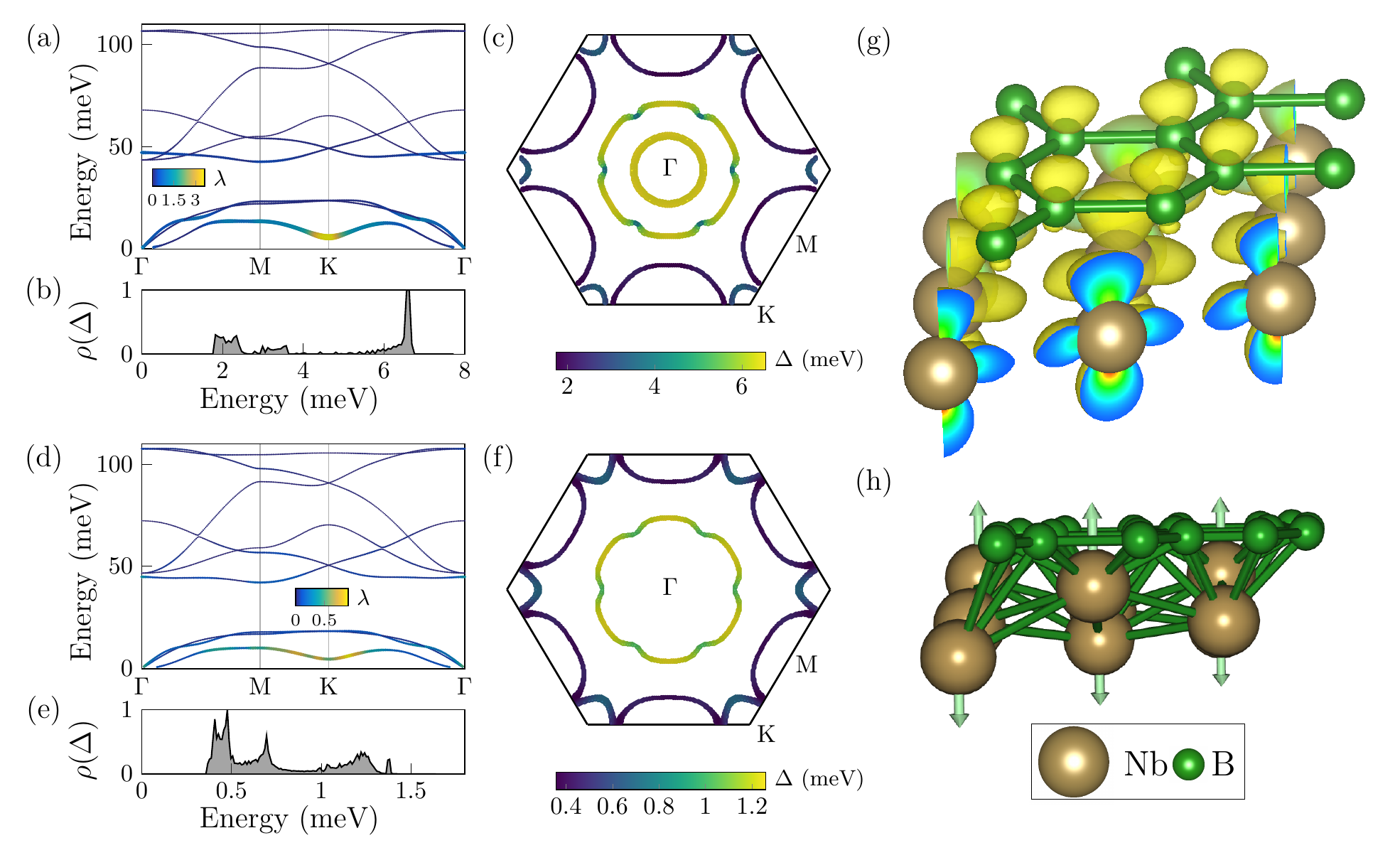}
\caption{Phonon dispersion with the mode-resolved electron-phonon coupling (EPC) indicated by colors and thickness for (a) NbB$_2$ and (d) TaB$_2$. Distribution of the superconducting gap of (b) NbB$_2$ at 5 K and (e) TaB$_2$ at 2 K. Momentum-dependent superconducting gap on the Fermi surface for (c) NbB$_2$ and at 5 K (f) TaB$_2$ at 2 K. (g) Norm of the wavefunction of the $d_{z^2}$ state from the circular sheets around the $\Gamma$ point of NbB$_2$. (h) Vibrational mode corresponding to the softening in the lowest acoustic phonon modes of NbB$_2$ and TaB$_2$. }
\label{fig2}
\end{figure*}

Among these materials, CaB$_4$, CaB$_2$, NbB$_2$, and ScB$_2$ with high $T_c$ values 36.1, 41.6, 35.5, and 20.4~K were identified as the most prominent superconductors. Also, the group-5-based compounds VB$_2$, NbB$_2$, TaB$_2$ are worthy of further investigation. CaB$_2$ is similar to MgB$_2$ in terms of electronic properties, as one can see in Supplementary Material. Therefore, two distinct sets of partially occupied bands contribute to superconductivity, mainly arising from covalent bonding, $\sigma$ ($s$ + $p_{x,y}$) and non-bonding $\pi$ ($p_z$), similar to MgB$_2$~\cite{PhysRevLett.86.4366}. In addition to the $\sigma$ states around the $\Gamma$ point, the non-bonding $\pi$ orbitals, contributing Fermi level states centered around the K point, have distinct superconducting properties, as is also the case for MgB$_{2}$. As might be foreseen, the strong hybridization between B $2p$ and transition metal $d$ states should be taken into consideration for VB$_2$, NbB$_2$, TaB$_2$, and ScB$_2$. Zhang \textit{et al.} investigated how $d$ orbitals stemming from Nb split into ($d_{xy}$, $d_x-d_y$), ($d_{xz}$,$d_{yz}$), and ($d_{z^2}$) around the Fermi level in NbB$_2$~\cite{doi:10.1021/acs.jpclett.9b00762}. 

Having evaluated the results obtained within the isotropic limit, we decided to focus further on several materials with the most promising superconducting properties. Thus, we performed fully anisotropic Eliashberg calculations for CaB$_2$, CaB$_4$, VB$_2$, NbB$_2$, TaB$_2$, and ScB$_2$ in order to fully characterize the origin and behavior of superconductivity in MBenes.

Calcium borides -- CaB$_2$ and CaB$_4$ -- have proven to be the most prominent superconductors among the considered structures. The Fermi surface of CaB$_2$, shown in Fig. \ref{fig1} (c), closely resembles that of MgB$_2$ and consists of two $\sigma$ bands, a $\pi$ band and a surface band. Both $\sigma$ and $\pi$ electronic states are localized in the layer of boron atoms while the surface state is mainly localized on top of the Ca layer facing vacuum. The Fermi surface of CaB$_4$, shown in Fig. \ref{fig1} (f), for the most part retains the same features as CaB$_2$ with one exception. Since the calcium layer in CaB$_4$ is fully encapsulated between two boron layers, this surface state is suppressed. The phonon band structures of the calcium borides, as shown in Fig. \ref{fig1} (a) and (d), are quite different. In CaB$_2$, acoustic and optical phonon modes are separated by an energy gap, which gets completely suppressed in CaB$_4$, notwithstanding the similarity of the corresponding atomic displacements in both compounds. These different phonon dispersions lead to different electron-phonon coupling (EPC) in the two compounds, as shown in Fig. \ref{fig1} (a) and (d). As can be seen from the Eliashberg function $\alpha^2F$ of CaB$_2$, acoustic and optical phonon modes contribute almost equally to the total EPC of 1.59, whereas for CaB$_4$, due to the absence of the gap in the phonon dispersion, one cannot make a clear distinction in the Eliashberg function between acoustic and optical phonon modes. The total EPC of the calcium tetraboride is 1.43 which is slightly lower than in the diboride. As in the case of MgB$_2$, in both calcium boride compounds, a significant part of the total EPC comes from one in-plane optical phonon mode, belonging to the $E_{2}$ class. $\lambda_{\mathrm{mode}}$ also reveals differences in the character of the EPC in CaB$_2$ and CaB$_4$. Highly coupling phonon states in the former are fairly localized around the $\Gamma$ point while in the latter there is also a sizable contribution from other parts of the Brillouin zone (particularly along $\Gamma$-M).

Based on our fully anisotropic Eliashberg calculations, we can analyze how different electronic states on the Fermi surface contribute to superconductivity, through the anisotropic gap spectrum $\Delta(\mathbf{k})$. At a temperature of 5 K, the largest $\Delta$ values, reaching as high as 17.4 meV in CaB$_2$, stem from the B-$\sigma$ states, while gaps of the B-$\pi$ and surface states are highly anisotropic, ranging from 1.7 to 7.5 meV (Fig. \ref{fig1} (c)). The superconducting gap distribution for calcium diboride, as shown in Fig. \ref{fig1} (b), shows two distinct superconducting gaps: the strongest $\sigma$-state gap, and a highly anisotropic gap which is a hybrid formed by the $\pi$ and the surface electronic states. We proceeded solving the anisotropic Eliashberg equations for increasing temperatures until the gaps vanished, yielding a critical temperature of CaB$_2$ of around 70 K. The picture for CaB$_4$ is overall quite similar, with some notable differences. Since the surface electronic state is absent in the tetraboride, its lowest gap is now only formed by the $\pi$ electronic state, and thus is much less anisotropic (Fig. \ref{fig1} (e)). At 5 K temperature, the highest gap value of CaB$_4$ (15.6 meV) is less strong than in CaB$_2$. Moreover, the two $\sigma$ electronic sheets on the Fermi surface, which form the strongest gap in the diboride case, give rise to two slightly separated gaps in the tetraboride case (by less than 1 meV at 5 K). As compared to CaB$_4$, the slightly weaker gap energies of CaB$_4$ result in a lower critical temperature of 64 K.
\begin{figure*}[!ht]
\includegraphics[width=\linewidth]{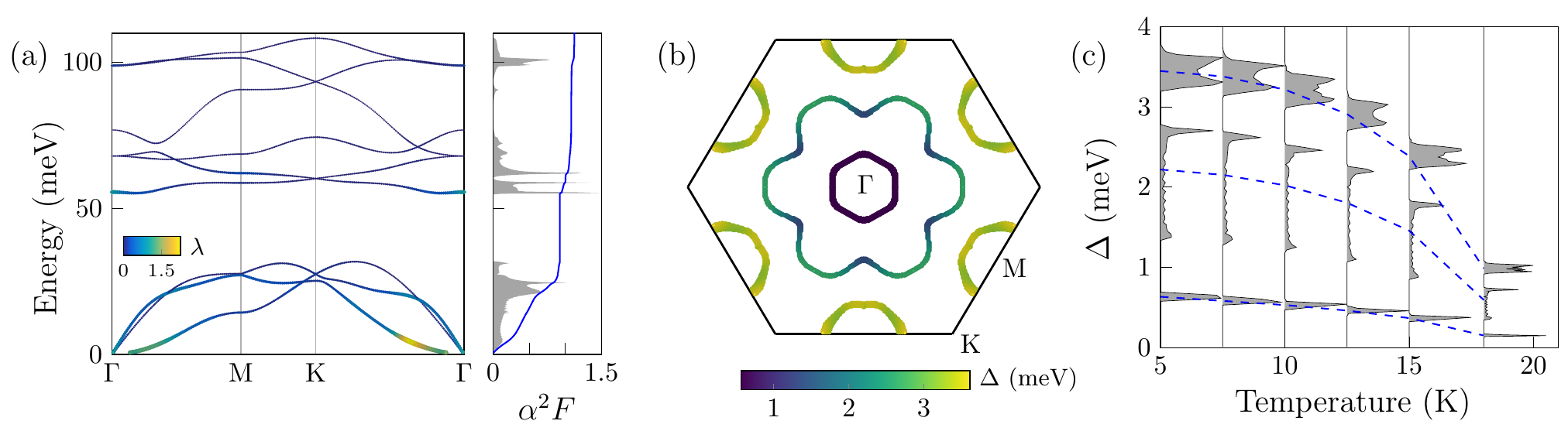}
\caption{Superconducting properties of VB$_2$: (a) Phonon dispersion with the mode-resolved electron-phonon coupling (EPC) indicated by colors and size. (b) Momentum-dependent superconducting gap on the Fermi surface. (c) Evolution of the superconducting gap distribution with temperature.}
\label{fig3}
\end{figure*}

Diborides of transition metals from group 5 of the periodic table -- V, Nb and Ta -- also show promise to be good superconductors. Particularly, the critical temperature of NbB$_2$ in the isotropic limit is 35.5 K, which is 3 times higher than the isotropic value in monolayer MgB$_2$. However, for Nb and Ta diborides, we found that electronic smearing has a critical influence on phonons and EPC. With the standard electronic smearing value of 0.01 Ha, both structures show negative phonon frequencies in the vicinity of the K point, suggesting that the structures are not in their ground state. The unstable phonon mode, as shown in Fig. \ref{fig2} (h), is an acoustic mode with out-of-plane vibrations of Nb/Ta atoms. 

By increasing the smearing value to 0.03 Ha, we managed to get rid of the negative phonons (Fig. \ref{fig2} (a) and (d)). However, the instability transforms into a softening in the phonon dispersions of both compounds at the K point. These dips, as shown in Fig.~\ref{fig2} (a) and (d), form EPC hotspots with the values of the coupling much higher compared to all other phonon modes. Thus, the total EPC and the critical temperature of NbB$_2$ and TaB$_2$ strongly depend on the electronic smearing used in the calculation and so cannot be assessed precisely. It is nevertheless interesting that both compounds, despite being quite close in their electronic and vibrational properties, show quite disparate superconducting behavior. The superconducting gap structure is more complex in transition metal borides as compared to alkaline earth metal borides like MgB$_2$ and CaB$_2$. For NbB$_2$, this is mainly due to the outer semi-circular sheet around the $\Gamma$ point, shown in Fig. \ref{fig2} (c). The orbital character of this electronic sheet is not constant along the sheet because the character changes strongly at the points where the sheet shows indents. Namely, everywhere on this Fermi sheet, except for the indents, the electronic states mainly stem from the $d_{z^2}$ orbital of Nb and $p_z$ of B, which is similar to the states on the inner circular sheet around the $\Gamma$ point (Fig. \ref{fig2} (g)). These states from the two circular Fermi sheets form the strongest superconducting gap in NbB$_2$ (Fig. \ref{fig2} (b)). All the remaining Fermi sheets together with the indent states of the outer sheet around $\Gamma$ form the weakest, strongly anisotropic gap. Thus, in NbB$_2$, electronic states from one Fermi sheet contribute to two distinct superconducting gaps. TaB$_2$, calculated with the same parameters as NbB$_2$, shows very weak EPC compared to the former, and all the Fermi sheets form a single anisotropic gap (Fig. \ref{fig2} (e) and (f)). 

\begin{figure*}[!ht]
\includegraphics[width=\linewidth]{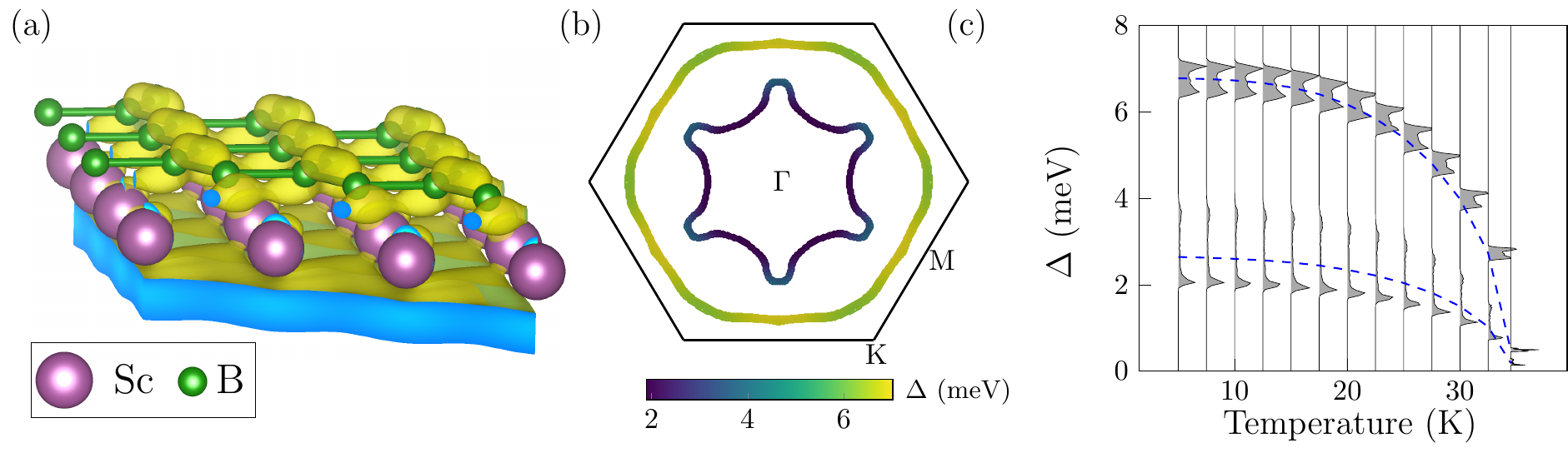}
\caption{(a) Surface electronic state of ScB$_2$, corresponding to the inner sheet of the Fermi surface. Superconducting properties of ScB$_2$, namely (b) momentum-dependent superconducting gap on the Fermi surface, and (c) evolution of the superconducting gap distribution with temperature.}
\label{fig4}
\end{figure*}

The lightest transition metal of group 5, vanadium, forms a diboride structure without unstable phonons, even when calculated with the electronic smearing as low as 0.01 Ha (Fig. \ref{fig3} (a)). Considering the out-of-plane nature of the unstable mode in both NbB$_2$ and TaB$_2$ (Fig. \ref{fig2} (h)), it appears that V, being lighter than its group neighbors, does not cause rippling of the structure, rendering it dynamically stable. Even though V, Nb and Ta belong to the same group, the Fermi surface of VB$_2$ (Fig. \ref{fig3} (b)) is quite different as compared to the diborides of Nb and Ta. In VB$_2$, there is no sheet around the K point and the two sheets around the $\Gamma$ point have a hexagonal and a flower-like shape respectively, in contrast to almost completely circular sheets in the two other diborides. The hexagonal sheet stems from a combination of out-of-plane \textit{d} orbitals with considerable contribution of the $d_{z^{2}}$ orbital, whereas the flower-shaped sheet is a mixture of in-plane and out-of-plane \textit{d} orbitals of V without $d_{z^{2}}$ contribution. VB$_2$ does share a common sheet with other diborides, namely the circular sheet around the M point which corresponds to out-of-plane \textit{d} orbitals of the M atom ($d_{xz}$ and $d_{yz}$).

The total electron-phonon coupling for VB$_2$ is 1.13 with the main contributions coming from the acoustic phonon modes as well as the lowest optical mode representing out-of-plane vibrations of boron atoms, as shown in Fig. \ref{fig3} (a). Interestingly, the superconducting gap on the Fermi surface shows the opposite pattern for VB$_2$ as compared to other MBenes of group 5. The electronic sheets around the $\Gamma$ point, which are attributed mainly to $d_{z^{2}}$ orbital of the M atom, represent the strongest gap in both NbB$_2$ and TaB$_2$ (Fig. \ref{fig2} (c) and (f)) while in VB$_2$ this state gives rise to the weakest gap. The is true for the circular state around the M point. Unlike in the case of Nb and Ta, VB$_2$ has an additional intermediate gap stemming from the flower-shaped Fermi sheet around the $\Gamma$ point. Thus, VB$_2$ is a three-gap superconductor. Our calculations show that the gaps vanish at the critical temperature of around 18.5 K. 

The last material we have considered in more detail was ScB$_2$. Scandium belongs to group 3 of the periodic table, on the border between alkaline earth metals and transition metals. Since Sc has only one \textit{d} electron in its outer shell, other electronic states can contribute more to the electronic states around the Fermi level. Therefore, ScB$_2$ represents an intermediate case between CaB$_2$ and the MBenes based on group 5 transition metals. The Fermi surface of ScB$_2$, shown in Fig. \ref{fig4} (b), consists of two sheets. The inner sheet corresponds to a surface state (Fig. \ref{fig4} (a)), akin to the one found in alkaline earth metal MBenes such as MgB$_2$ and CaB$_2$. However, in alkaline earth MBenes the surface state stems from Mg/Ca-$p$ states, whereas in ScB$_2$ it mainly has Sc-\textit{s} character, because of scandium's different electronic configuration. The states of the outer sheet are a mixture of the boron $p_{z}$ orbital and in-plane $d$ orbitals of Sc. The total isotropic EPC of ScB$_2$ amounts to 1.27. Anisotropic Eliashberg calculation show that ScB$_2$ is a two-gap superconductor (Fig. \ref{fig4} (b)), with the strongest gap stemming from the outer Fermi sheet and the weakest gap stemming from the inner sheet. The resulting critical temperature is rather high -- around 34.5 K. 

\section{Conclusion}

Our systematic investigation performed with both isotropic and anisotropic Eliashberg theory calculations broadly sheds light on the potential of two-dimensional metal boride systems as superconducting materials. High critical temperatures (up to 72 K), as well as multigap superconducting behavior in CaB$_2$, CaB$_4$, VB$_2$, NbB$_2$, TaB$_2$, and ScB$_2$ structures were shown and detailed in this work. Coupled with the advent of their fabrication \cite{C9TA08820B, YANG2021}, here shown different electronic structures and Fermi surface configurations together with the demonstrated richness of the superconducting behavior in 2D metal-borides recommend their versatility for further exploration in fundamental (multi)functional heterostructures as well as potential device applications.

\begin{acknowledgements}
This work is supported by Research Foundation-Flanders (FWO-Vlaanderen), Special Research Funds of the University of Antwerp (TOPBOF), The Scientific and Technological Research Council of Turkey (TUBITAK) under the contract number COST-118F187, and the Air Force Office of Scientific Research under award number FA9550-19-1-7048. Computational resources were provided by the High Performance and Grid Computing Center (TRGrid e-Infrastructure) of TUBITAK ULAKBIM and by the VSC (Flemish Supercomputer Center), funded by the FWO and the Flemish Government -- department EWI. J.B. is a postdoctoral fellow of the FWO. The collaborative effort in this work was supported by the EU COST Action CA16218 NANOCOHYBRI.
\end{acknowledgements}

\providecommand{\noopsort}[1]{}\providecommand{\singleletter}[1]{#1}%

\end{document}